\documentclass[11pt,twoside]{article}
\usepackage{asp2010}

\resetcounters

\begin{document}

\title{Detection of Periodic Variability in Simulated QSO Light Curves}
\author{David B. Westman$^1$, Chelsea L. MacLeod$^1$, and \v{Z}eljko Ivezi\'{c}$^{1,2}$
\affil{$^1$Department of Astronomy, Box 351580, University of Washington,\\ Seattle, WA 98195-1580 USA \\
 $^2$ University of Zagreb, Department of Physics, Bijeni\v{c}ka c.~32, P.P. 331, Zagreb, Croatia}}

\begin{abstract}
Periodic light curve behavior predicted for some binary black hole systems might be detected 
in large samples, such as the multi-million quasar sample expected from the Large Synoptic 
Survey Telescope (LSST). We investigate the false-alarm probability for the discovery of a periodic
signal in light curves simulated using damped random walk (DRW) model. This model provides 
a good description of observed light curves, and does not include periodic behavior. 
We used the Lomb-Scargle periodogram to search for a periodic signal in a million simulated light 
curves that properly sample the DRW parameter space, and the LSST cadence space. We find 
that even a very conservative threshold for the false-alarm probability still yields thousands
of ``good'' binary black hole candidates. We conclude that the future claims for binary black 
holes based on Lomb-Scargle analysis of LSST light curves will have to be interpreted with caution.
\end{abstract}

\section{Introduction}

Modern surveys of the sky, such as the Sloan Digital Sky Survey (SDSS, \citealt{SDSSoverview}), have 
collected huge amounts of data (20 TB for SDSS), requiring the development of automated analysis 
methods. The Large Synoptic Survey Telescope (LSST, \citealt{LSSToverview}) will gather even more 
data than the SDSS did (one SDSS equivalent per night over ten years of operations). Among other 
populations, LSST will identify several million quasars (QSO) and obtain their light curves. In this 
contribution, we discuss an automated analysis of a million simulated light curves to search for periodic variability. 

The optical variability of QSOs has been recognized since they were first identified \citep{MS63} and is aperiodic and on the 
order of 20\% on timescales of months to years \citep[for recent results see, e.g., ][hereafter M10]{macleod_2010}.
Periodic variability has been suggested as one of the observational characteristics of a binary black hole system \citep{komossa_2003},
\citep[also see][for an investigation of broad line emission spectra in binary black holes]{shen_2010},
but there is no convincing observational evidence for such systems yet.

\section{Goal}

The large QSO sample expected from LSST might enable a detection of periodic signal 
in observed light curves. Recently we employed the Lomb-Scargle periodogram \citep{lomb_1976,scargle_1982,horne_1986}
to test $\thicksim$9000 spectroscopically confirmed QSOs from SDSS Stripe 82 (S82) for periodic variability (see Appendix of M10). 
We reuse the tools developed for the analysis of SDSS data and study light curves simulated using a mathematical variability  model
trained on SDSS data, and sampled using simulated LSST cadences. 

\section{Creation of Light Curves}

Approximately $1x10^{6}$ QSO light curves were generated using the damped random walk model \citep[M10]{kelly_2009, kozlowski_2010}.
The difference between this model and the well-known random walk is that an additional self-correcting term pushes any deviations back
towards the mean flux on a time scale $\tau$. The above studies have  established that DRW can statistically explain the observed light curves 
of quasars at an impressive fidelity level (0.01-0.02 mag).

The input parameters to the model are the characteristic time scale, $\tau$, and the root-mean-square (rms) variability on long time scales, or structure function, \emph{sf}.  The input parameters were determined using the scalings with black hole mass ($M_{BH}$), absolute magnitude ($M_{i}$), and redshift found by M10.  These physical parameters were drawn from the distribution shown in Figure 1.  After generating $\thicksim$83,500 well-sampled light curves, each light curve was resampled to the 12 different simulated $r$-band LSST cadences from \citet[$\thicksim$200 observations spread over 10 years]{delgado_2006} to obtain $\thicksim\!1x10^{6}$ total light curves.  

\articlefigure[width=3.5in]{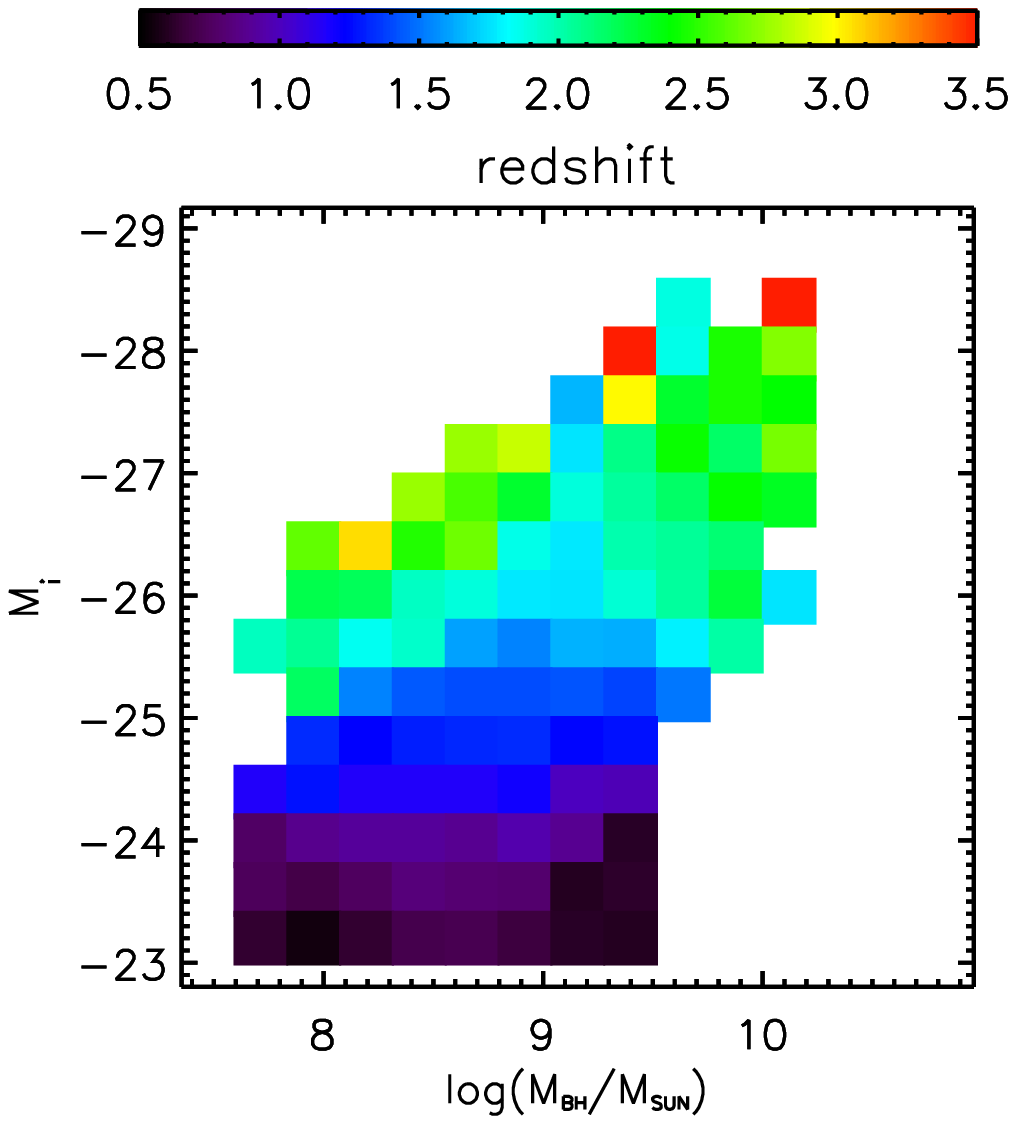}{Figure 1}{QSO model parameter ranges used to derive inputs to light-curve generation}
All light curves were analyzed with Lomb-Scargle periodogram software.  If the maximum power spectral density (PSD)
value was above the level set according to the false alarm probability, \emph{fap}, then the curve became part of a set used for further examination.

\section{Results}
We found that for high \emph{fap} values, the actual number of light curves which exceeded the \emph{fap} level was less than the theoretical value
by an appreciable amount.  For a \emph{fap} value of 5\%, there were 13,294 (1.3\%) light curves that exceeded that level, for a \emph{fap} value of 1\%,
there were 4696 (0.47\%), and for a fap value of 0.1\%, there were 1035 (0.1\%).

\articlefigure[width=4.0in,height=2.3in]{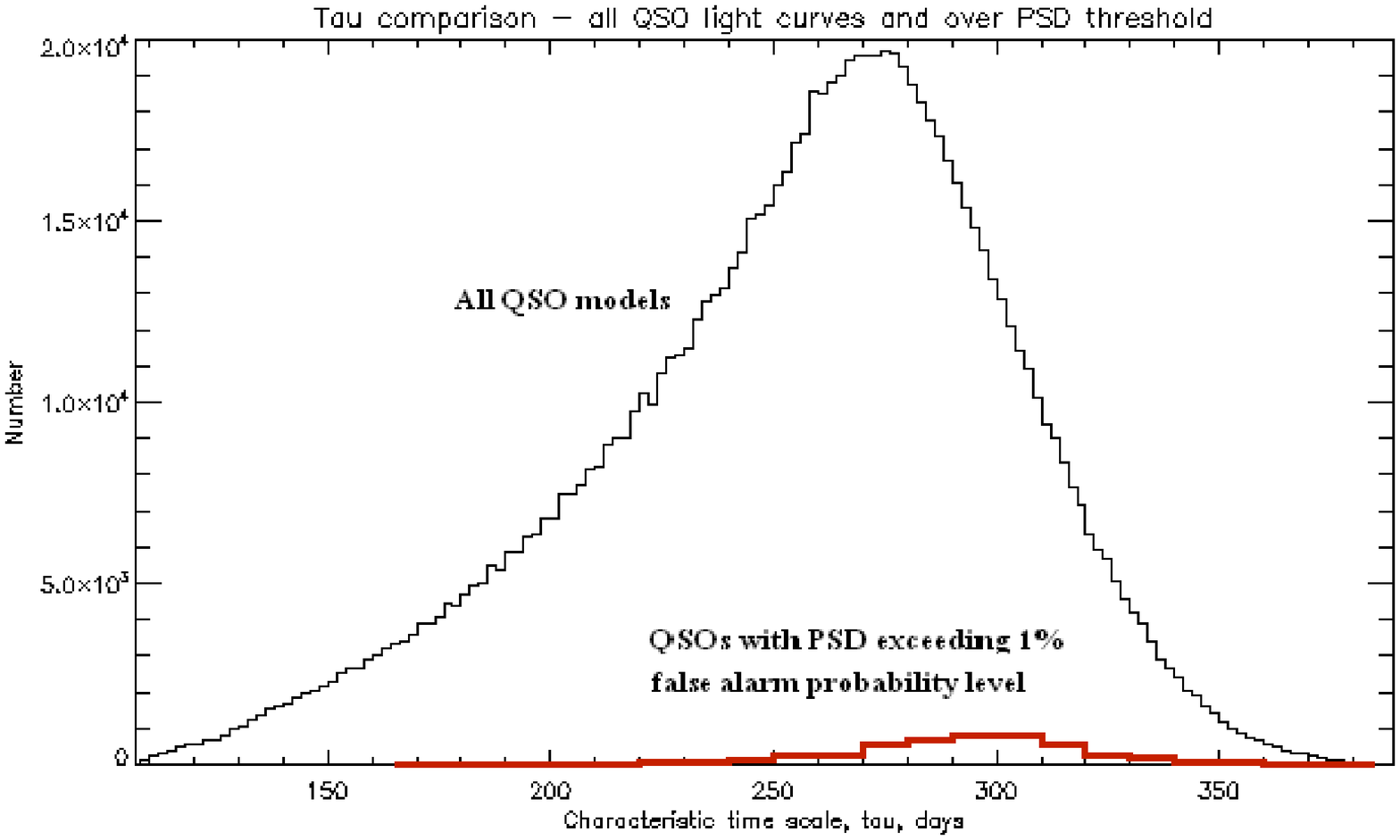}{Figure 2}{Distribution of characteristic time, $\tau$, for 
all QSO light curves compared with $\tau$ distribution for light curves with a peak SED value exceeding the 1\% \emph{fap} level}
\articlefigure[width=3.0in,height=2.0in]{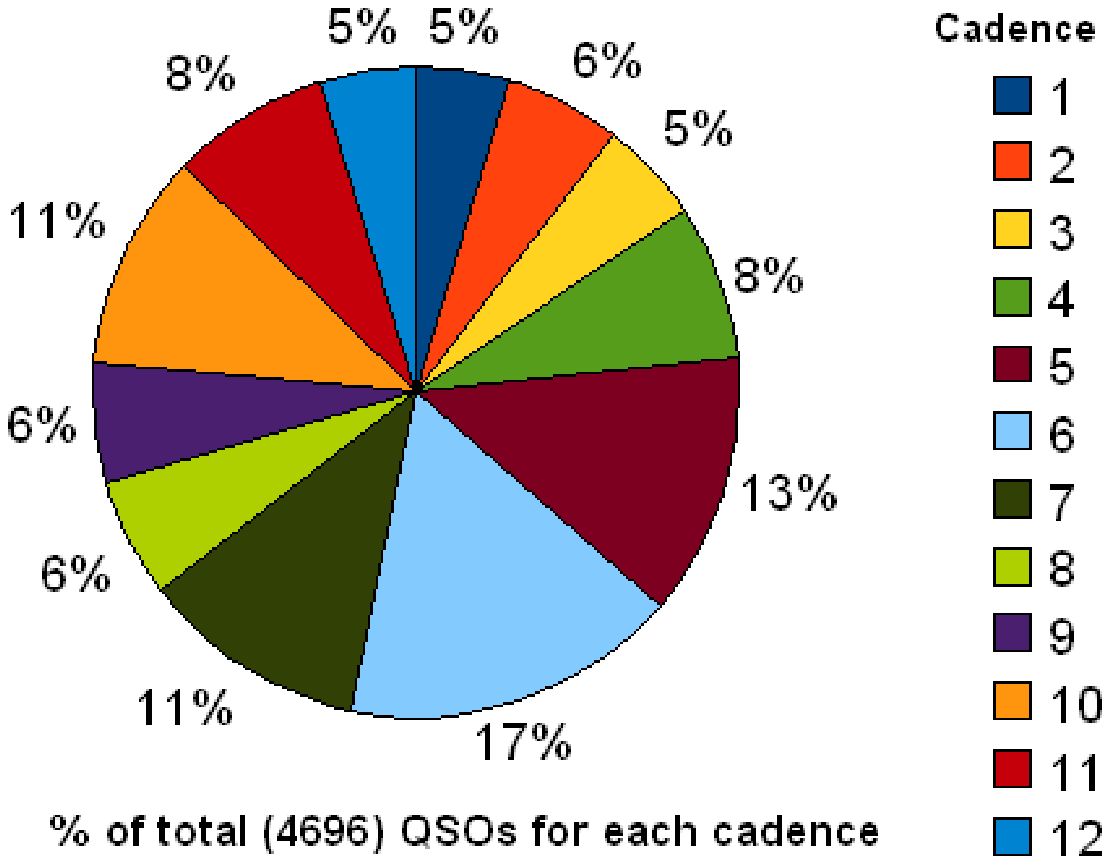}{Figure 3}{Proportion of QSOs with peak PSD values over 1\% \emph{fap} level sorted by LSST cadence}
Figure 2 shows that values of $\tau$ for QSO models with PSD exceeding the 1\% \emph{fap} level (red histogram) are distributed differently than those
for all the QSO models (black histogram).   This bias is due to the fact that when $\tau$ is long, only a few ``oscillations'' are observed over the
duration of the light curve, causing the periodogram to mistake the damped random walk behavior for a periodic behavior.  

Figure 3 compares the proportions of QSO models found to exceed the 1\% \emph{fap} level for each of the cadences.  There is a marked difference between
the various cadences here, showing that some of the cadences allow many more periodogram results for which the 1\% \emph{fap} level is exceeded.
This may be because the variations in the sampling pattern and the ``windowing'' effect of the Lomb-Scargle periodogram can cause a resonance effect at
some test frequencies.  Therefore, the results obtained by using the Lomb-Scargle periodogram method can be greatly influenced by the pattern of the
observations used.

\section{Conclusion}
This work shows that the Lomb-Scargle periodogram method may be useful for detecting potentially periodic behavior in QSO light curves in a large-scale surveys, such as the one to be carried out by the LSST.  However, even with \emph{fap} as small as 0.1\%, the large LSST sample would yield $\thicksim$1,000 \textbf{false} candidates.
Therefore, future claims of periodic behavior based on Lomb-Scargle analysis of LSST light curves will have to be interpreted with caution. In particular, black hole binary candidates identified by this method would have to be examined individually and with supplemental observations.

\acknowledgements We acknowledge support by NSF grant AST-0807500 to the University of Washington, and NSF grant AST-0551161
to the LSST for design and development activity. \v{Z}. Ivezi\'{c} thanks the University of Zagreb, where a portion of this work
was completed, for its hospitality, and acknowledges support by the Croatian National Science Foundation grant O-1548-2009.
\bibliography{P124}

\begin{thebibliography}{}
\expandafter\ifx\csname natexlab\endcsname\relax\def\natexlab#1{#1}\fi
\expandafter\ifx\csname url\endcsname\relax
  \def\url#1{\texttt{#1}}\fi
\expandafter\ifx\csname urlprefix\endcsname\relax\def\urlprefix{URL }\fi
\providecommand{\eprint}[2][]{\url{#2}}

\bibitem[{Delgado et~al.(2006)}]{delgado_2006}
Delgado, F., et~al. 2006, in Observatory Operations: Strategies, Processes, and
  Systems, edited by D.~R. Silva, \& R.~E. Doxsey (Bellingham, WA, USA: SPIE),
  vol. 6270 of Proc. SPIE, 62701D

\bibitem[{Horne \& Baliunas(1986)}]{horne_1986}
Horne, J.~H., \& Baliunas, S.~L. 1986, ApJ, 302, 757

\bibitem[{{Ivezi\'{c}} et~al.(2008){Ivezi\'{c}}, {Tyson}, {Allsman}, {Andrew},
  {Angel}, \& {for the LSST Collaboration}}]{LSSToverview}
{Ivezi\'{c}}, {\v{Z}}., {Tyson}, J.~A., {Allsman}, R., {Andrew}, J., {Angel},
  R., \& {for the LSST Collaboration} 2008, ArXiv e-prints.
  \eprint{arXiv:0805.2366}

\bibitem[{Kelly et~al.(2009)Kelly, Bechtold, \& Siemiginowska}]{kelly_2009}
Kelly, B.~C., Bechtold, J., \& Siemiginowska, A. 2009, ApJ, 698, 895

\bibitem[{Komossa(2003)}]{komossa_2003}
Komossa, S. 2003, in The Astrophysics of Gravitational Wave Sources, edited by
  J.~Centrella (Mellville, NY, USA: AIP), vol. 686 of AIP Conf. Proc., 161

\bibitem[{Kozlowski et~al.(2010)}]{kozlowski_2010}
Kozlowski, S., et~al. 2010, ApJ, 708, 927

\bibitem[{Lomb(1976)}]{lomb_1976}
Lomb, N.~R. 1976, Ap\&SS, 39, 447

\bibitem[{MacLeod et~al.(2010)}]{macleod_2010}
MacLeod, C.~L., et~al. 2010, ApJ, 721, 1014

\bibitem[{Matthews \& Sandage(1963)}]{MS63}
Matthews, T.~A., \& Sandage, A.~R. 1963, ApJ, 138, 30

\bibitem[{Scargle(1982)}]{scargle_1982}
Scargle, J.~D. 1982, ApJ, 263, 835

\bibitem[{Shen \& Loeb(2010)}]{shen_2010}
Shen, Y., \& Loeb, A. 2010, ApJ, 725, 249

\bibitem[{{York} et~al.(2000)}]{SDSSoverview}
{York}, D.~G., et~al. 2000, \aj, 120, 1579. \eprint{arXiv:astro-ph/0006396}

\end{thebibliography}

\end{document}